\documentclass[aps,prl,twocolumn,groupedaddress,superscriptaddress,citeautoscript,longbibliography,a4paper]{revtex4-2}
\usepackage{amsmath,amssymb,amsfonts,bm}
\usepackage{xcolor}
\usepackage{graphicx}
\usepackage{epstopdf}
\usepackage{dcolumn}
\usepackage{mathrsfs}
\usepackage{tgtermes}

\usepackage{hyperref}
\hypersetup{colorlinks,
	linkcolor={blue!75!black!80!yellow},
	citecolor={blue!75!black!80!yellow},
	urlcolor={blue!75!black!80!yellow}
}

\bibpunct{[}{]}{,}{n}{}{}
\usepackage{verbatim}
\usepackage{amsthm,amsmath,amssymb}
\usepackage{mathrsfs}
\usepackage{booktabs}
\usepackage{multirow}
\usepackage{pifont}
\makeatletter
\newcommand{\fmarki}{\#}
\newcommand{\fmarkii}{*}
\newcommand{\fmarkiii}{\ensuremath{\dagger}}
\newcommand{\fmarkiv}{\ensuremath{\ddagger}}
\newcommand{\fmarkv}{\ensuremath{\mathsection}}
\newcommand{\fmarkvi}{\ensuremath{\mathparagraph}}
\newcommand{\fmarkvii}{\ensuremath{\|}}
\newcommand{\fmarkviii}{**}

\def\@fnsymbol#1{{\ifcase#1\or \fmarki\or \fmarkii\or \fmarkiii\or \fmarkiv\or \fmarkv\or \fmarkvi\or \fmarkvii\or \fmarkviii\or \fmarkix \else\@ctrerr\fi}}
\makeatother

\begin{document}

\title{Acoustic Three-dimensional Chern Insulators with Arbitrary Chern Vectors}

%

\author{Linyun Yang}
\thanks{L. Y., X. X., Y. M. and Z. Z. contributed equally to this work.}
\affiliation{Department of Electrical and Electronic Engineering, Southern University of Science and Technology, Shenzhen 518055, China}
\author{Xiang Xi}
\thanks{L. Y., X. X., Y. M. and Z. Z. contributed equally to this work.}
\affiliation{Department of Electrical and Electronic Engineering, Southern University of Science and Technology, Shenzhen 518055, China}

\author{Yan Meng}
\thanks{L. Y., X. X., Y. M. and Z. Z. contributed equally to this work.}
\affiliation{Department of Electrical and Electronic Engineering, Southern University of Science and Technology, Shenzhen 518055, China}

\author{Zhenxiao Zhu}
\thanks{L. Y., X. X., Y. M. and Z. Z. contributed equally to this work.}
\affiliation{Department of Electrical and Electronic Engineering, Southern University of Science and Technology, Shenzhen 518055, China}

\author{Ying Wu}
\affiliation{School of Science, Nanjing University of Science and Technology, Nanjing, 210094, China}

\author{Jingming Chen}
\affiliation{Department of Electrical and Electronic Engineering, Southern University of Science and Technology, Shenzhen 518055, China}

\author{Minqi Cheng}
\affiliation{Department of Electrical and Electronic Engineering, Southern University of Science and Technology, Shenzhen 518055, China}

\author{Kexin Xiang}
\affiliation{Department of Electrical and Electronic Engineering, Southern University of Science and Technology, Shenzhen 518055, China}

\author{\\ Perry Ping Shum}
\affiliation{Department of Electrical and Electronic Engineering, Southern University of Science and Technology, Shenzhen 518055, China}

\author{Yihao Yang}
\affiliation{Interdisciplinary Center for Quantum Information, State Key Laboratory of Modern Optical Instrumentation, ZJU-Hangzhou Global Science and Technology Innovation Center, College of Information Science and Electronic Engineering, ZJU-UIUC Institute, Zhejiang University, 310027 Hangzhou, China}

\author{Hongsheng Chen}
\affiliation{Interdisciplinary Center for Quantum Information, State Key Laboratory of Modern Optical Instrumentation, ZJU-Hangzhou Global Science and Technology Innovation Center, College of Information Science and Electronic Engineering, ZJU-UIUC Institute, Zhejiang University, 310027 Hangzhou, China}

\author{Jian Li}
\affiliation{School of Science, Westlake University, 18 Shilongshan Road, Hangzhou 310024, Zhejiang Province, China}

\author{Bei Yan}
\email{yanb3@sustech.edu.cn}
\affiliation{Department of Electrical and Electronic Engineering, Southern University of Science and Technology, Shenzhen 518055, China}

\author{Gui-Geng Liu}
\email{guigeng001@e.ntu.edu.sg}
\affiliation{Division of Physics and Applied Physics, School of Physical and Mathematical Sciences, Nanyang Technological University, 21 Nanyang Link, Singapore 637371, Singapore}

\author{Baile Zhang}
\email{blzhang@ntu.edu.sg}
\affiliation{Division of Physics and Applied Physics, School of Physical and Mathematical Sciences, Nanyang Technological University, 21 Nanyang Link, Singapore 637371, Singapore}
\affiliation{Centre for Disruptive Photonic Technologies, The Photonics Institute, Nanyang Technological University, Singapore, Singapore}

\author{Zhen Gao}
\email{gaoz@sustech.edu.cn}
\affiliation{Department of Electrical and Electronic Engineering, Southern University of Science and Technology, Shenzhen 518055, China}


\begin{abstract}
	
The Chern vector is a vectorial generalization of the scalar Chern number, being able to characterize the topological phase of three-dimensional (3D) Chern insulators. Such a vectorial generalization extends the applicability of Chern-type bulk-boundary correspondence from one-dimensional (1D) edge states to two-dimensional (2D) surface states, whose unique features, such as forming nontrivial torus knots or links in the surface Brillouin zone, have been demonstrated recently in 3D photonic crystals. However, since it is still unclear how to achieve an arbitrary Chern vector, so far the surface-state torus knots or links can emerge, not on the surface of a single crystal as in other 3D topological phases, but only along an internal domain wall between two crystals with perpendicular Chern vectors. Here, we extend the 3D Chern insulator phase to acoustic crystals for sound waves, and propose a scheme to construct an arbitrary Chern vector that allows the emergence of surface-state torus knots or links on the surface of a single crystal. These results provide a complete picture of bulk-boundary correspondence for Chern vectors, and may find use in novel applications in topological acoustics.  

\end{abstract}

\maketitle

The topological invariant of Chern number, which used to characterize two-dimensional (2D) Chern insulators \cite{haldane_1988,chang_experimental_2013,deng_quantum_2020,zhao_tuning_2020,serlin_intrinsic_2020,haldane_2008,wang_reflection_2008,wang_observation_2009,poo_2011,skirlo_2014,skirlo_2015}, has been generalized from a scalar to a vector \cite{haldane_berry_2004,berry_2018} in the form of $\bm{C}=(c_1,c_2,c_3)$ that is able to characterize three-dimensional (3D) Chern insulators. A remarkable feature of the bulk-boundary correspondence arising from the vectorial nature of Chern vector is that topological surface states of 3D Chern insulators can form nontrivial torus knots or links in the surface Brillouin zone (BZ), as recently demonstrated in 3D photonic crystals \cite{liu_topo_2022,devescovi_cubic_2021,devescovi_vector_2022}. However, unlike in other 3D topological phases, so far the nontrivial surface-state torus knots or links can emerge only along an internal domain wall between two 3D Chern insulators with perpendicular Chern vectors \cite{liu_topo_2022,devescovi_cubic_2021,devescovi_vector_2022}. This apparently is not the full picture of bulk-boundary correspondence for the Chern vector in 3D Chern insulators. 

Along a separate line of development, the acoustic analogue of Chern insulators known as acoustic Chern insulators \cite{yang_topo_prl_2015,khanikaev_topo_2015,ni_topo_2015,ding_exp_2019} has stimulated the rise of research in topological acoustics \cite{hecheng_2016,fleury_floquet_2016,lu_2017,zhangTopologicalSound2018,ma_topological_2019,he_acoustic_2020,jia_tunable_2021,xue_topological_2022,zhang_second_2023}, with promising sound manipulation with robust chiral edge states that are completely immune to backscattering from defects and disorders. However, despite lots of recent progress, the acoustic Chern insulators have so far been limited to 2D. 

In this Letter, we extend the concept of Chern vector to acoustic crystals, and propose a scheme to construct an acoustic 3D Chern insulator. In particular, the acoustic 3D Chern insulator can exhibit an arbitrary Chern vector, which allows the emergence of torus knots or links on the surface of a single crystal. Moreover, we also show the phenomenon of successive topological negative refraction at three adjacent facets of a single crystal, in which acoustic chiral surface states can pass through sharp edges continuously in 3D space through negative refraction without backscattering, being different from previously proposed topological negative refraction \cite{he_topological_2018,yang_topological_2019}.

\begin{figure*}[t]
	\centering
	\includegraphics{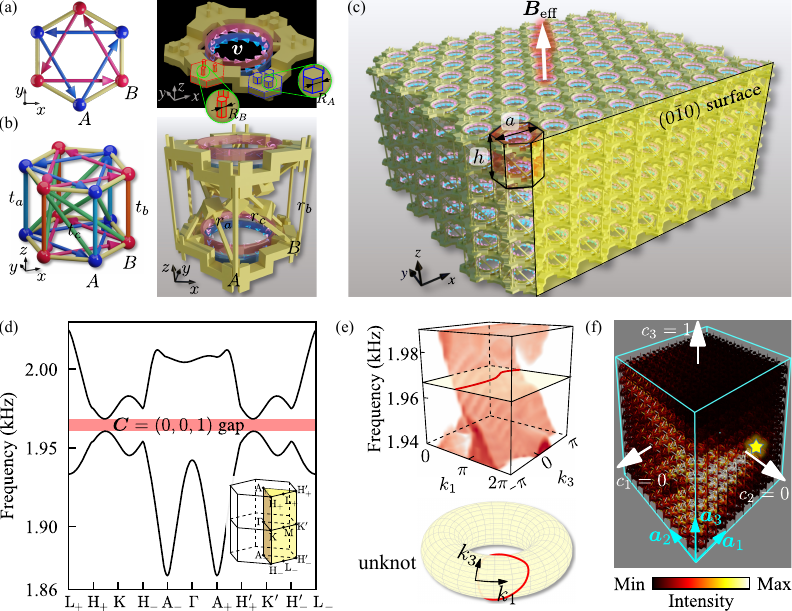}
	\caption{Acoustic 3D Chern insulator with a Chern vector $\bm{C}=(0,0,1)$. The tight-binding configuration and the acoustic crystal design of (a) a 2D Haldane model and (b) its 3D generalization. The red and blue arrows in the right panels indicate the circulating airflow among sites in sublattice $A$ and sublattice $B$, respectively. (c) Acoustic realization of a 3D Haldane model constructed by the unit cell shown in the right panel of (b), which exhibits an effective magnetic field $\bm{B}_{\mathrm{eff}}$ similar to the external magnetic field in previous 3D photonic Chern insulators. (d) The bulk band diagram of the acoustic 3D Chern insulator with a nontrivial topological $\bm{C}=(0,0,1)$ bandgap. (e) The $(0\bar{1}0)$-surface dispersion of the acoustic 3D Chern insulator and the unknot formed by the Fermi loop. (f) Full-wave simulated acoustic intensity distributions of the chiral surface states excited by a point source (yellow star) at the mid-gap frequency 1.968 kHz.
	}
	\label{fig:Fig1}
\end{figure*}

We start with the celebrated 2D Haldane model \cite{haldane_1988} and its acoustic realization -- the acoustic 2D Chern insulator, as shown in Fig. \ref{fig:Fig1}(a). The hexagonal prisms with three attached pillars represent the sublattice sites, whose on-site potentials can be tuned by changing the side widths of the pillars denoted as $R_A$ and $R_B$ [Fig. \ref{fig:Fig1}(a)]. The on-site potential difference $\Delta R=(R_A-R_B)/2$ represents the inversion symmetry (IS) breaking strength. In addition, the sites within each sublattice (Sites $A$ or Sites $B$) are coupled by a circulating airflow (small red and blue arrows) to implement the complex next-nearest-neighboring (NNN) hopping, and the circumferential speed $v$ of the flow characterizes the time-reversal symmetry (TRS) breaking strength \cite{yang_topo_prl_2015,khanikaev_topo_2015,ni_topo_2015,ding_exp_2019,fleury_sound_2014}. The two sublattices are slightly shifted vertically to avoid self-intersecting, and the real nearest-neighboring (NN) hopping between neighboring sites are implemented by Z-shaped air tubes (see details in the Supplemental Material \cite{supp}). Note that such TRS breaking by airflow has been utilized to realize an acoustic 2D Chern insulator recently \cite{ding_exp_2019}.

By stacking the acoustic 2D Haldane model periodically along the vertical direction with vertical and chiral interlayer couplings, we can construct an acoustic realization of 3D Haldane model with the effective magnetic field $\bm{B}_{\mathrm{eff}}$ parallel to the $z$-axis, as shown in Figs. \ref{fig:Fig1}(b) and \ref{fig:Fig1}(c). The vertical $(t_a,t_b)$ and chiral $(t_c)$ interlayer couplings are controlled by the side widths of the corresponding air tubes denoted as $r_a,r_b$ and $r_c$, as shown in Fig. 1(b). The chiral hopping is introduced to guarantee the emergence of ideal type-I Weyl points, facilitating the opening of a 3D complete bandgap with a relatively small TRS-breaking strength \cite{supp}. When varying the TRS ($v$) and IS ($\Delta R$) breaking strengths, the calculated phase diagram of the acoustic 3D Haldane model exhibits three different topological phases including trivial insulators, Weyl semimetals, and 3D Chern insulators (see details in the Supplemental Material \cite{supp}). Figure \ref{fig:Fig1}(d) shows the bulk band structure of an acoustic 3D Chern insulator with $\Delta R=0$ and $v=6~\mathrm{m/s}$, which opens a 3D topological bandgap with a nontrivial Chern vector $\bm{C}=(0,0,1)$. To gain more insights into the topological surface states of the acoustic 3D Chern insulator, Fig. \ref{fig:Fig1}(e) presents the front $(0\bar{1}0)$ surface density of states (DOS) \cite{sha_surface_2021} of the acoustic 3D Chern insulator, in which the solid red curve represents the Fermi loop and $k_i=\bm{k}\cdot \bm{a}_i \in [0,2\pi]$ denotes the reduced wave number. The three primitive lattice vectors are $\bm{a}_1=(a,0,0),\bm{a}_2=(a/2,\sqrt{3}a/2,0)$ and $\bm{a}_3=(0,0,h)$, where $a$ and $h$ denote the in-plane and out-of-plane lattice constants, respectively. It can be seen that the chiral surface states always exist for any $k_3$ and span the whole bandgap with negative group velocity. Since the Chern vector has only a single nonzero Chern number $c_3$, the Fermi loop winds a cycle only along $k_3$ direction but not $k_1$ direction, forming a simple loop (i.e., the unknot) on the 2D surface BZ. We then perform full-wave simulations to demonstrate the one-way and robust transport of the acoustic chiral surface states. As shown in Fig. \ref{fig:Fig1}(f), a point source (yellow star) is placed at the center of $(0\bar{1}0)$ surface. The excited acoustic chiral surface states propagate nonreciprocally and route around two sharp corners without notable reflection. Even if we insert a solid defect in the path of the chiral surface states, they still can bypass the obstacle with negligible reflection, revealing the topological protection of the acoustic chiral surface states (see details in the Supplemental Material \cite{supp}).

\begin{figure*}[t]
	\centering
	\includegraphics{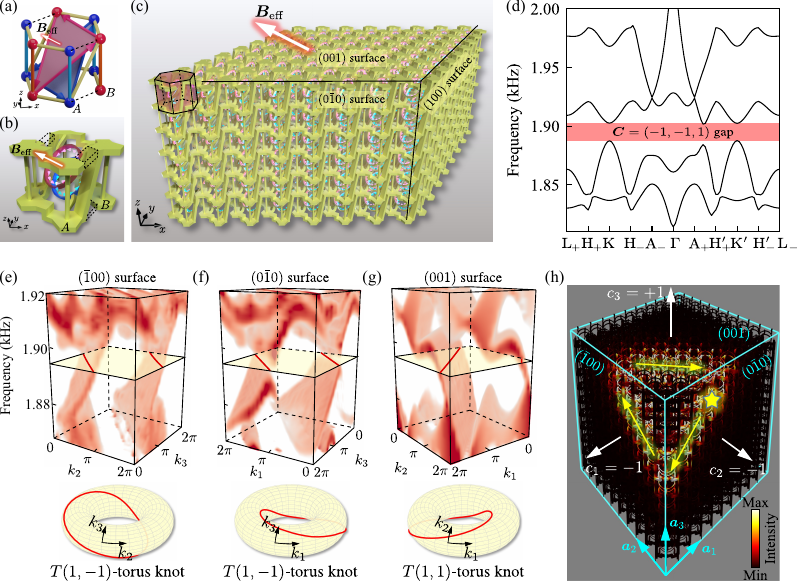}
	\caption{Acoustic 3D Chern insulator with a Chern vector $\bm{C}=(-1,-1,1)$. (a) The tilted 3D Haldane model and (b) its acoustic realization, in which the circulating airflows (blue and red arrows) are tilted to induce nonreciprocal hoppings between adjacent layers. (c) Schematic view of the acoustic 3D Chern insulator with a Chern vector $\bm{C}=(-1,-1,1)$. (d) The simulated bulk band structure of the acoustic 3D Chern insulator with a 3D topological $\bm{C}=(-1,-1,1)$ bandgap. (e)-(g) The chiral surface states dispersions (surface DOS) of the left $(\bar{1}00)$, front $(0\bar{1}0)$ and top $(001)$ surfaces, respectively. The lower panels show the torus knots formed by winding the Fermi loops on three different surface BZs. (h) Full-wave simulated acoustic intensity distributions of the chiral surface states excited by a point source (yellow star) at 1.897 kHz on the surfaces of an acoustic 3D Chern insulator with a Chern vector $\bm{C}=(-1,-1,1)$.
	}
	\label{fig:Fig2}
\end{figure*}

Note that the Chern vector $\bm{C}=(0,0,c_3)$ of the acoustic 3D Chern insulators discussed above has only a single nonzero Chern number $c_3=1$. In this case, only the four side surfaces parallel to the Chern vector can support chiral surface states, while the top and bottom surfaces cannot. Next, we demonstrate that, by tilting the circulating airflows and the effective magnetic fields, we can construct an acoustic 3D Chern insulator with a Chern vector $\bm{C}=(-1,-1,1)$ that supports chiral surface states on all six surfaces. These chiral surface states can form reflectionless topological negative refraction at the one-dimensional hinges separating different facets. As shown in Figs. \ref{fig:Fig2}(a)-\ref{fig:Fig2}(c), all circulating airflows (small blue and red arrows) are tilted and each of them couples three sublattice sites residing in adjacent layers (see details in the Supplemental Material \cite{supp}). Consequently, the effective magnetic field $\bm{B}_{\mathrm{eff}}$ is tilted with nonzero components along all primitive lattice vectors $\bm{a}_i$. Meanwhile, four NN hoppings (black dashed lines) are replaced by two slanted interlayer couplings (tilted yellow tubes) between different sublattice sites, tilting the 3D honeycomb lattice in according with the tilted effective magnetic fields $\bm{B}_{\mathrm{eff}}$. Such a tilted 3D Haldane model can be approximately described by a new Hamiltonian modified from the previously discussed 3D Haldane model and the Wilson-loop calculation of the tilted 3D Haldane model exhibits a 3D Chern insulating phase with a Chern vector $\bm{C}=(-1,-1,1)$  \cite{supp}. Figure \ref{fig:Fig2}(d) shows the simulated bulk band structure of the acoustic 3D Chern insulator, which exhibits a complete nontrivial topological Chern bandgap with a Chern vector $\bm{C}=(-1,-1,1)$. 

Compared with the acoustic 3D Chern insulator with a Chern vector $\bm{C}=(0,0,1)$, the acoustic 3D Chern insulator with a Chern vector $\bm{C}=(-1,-1,1)$ exhibits two unique features. First, the latter supports chiral surface states on all surfaces, as shown in Figs. \ref{fig:Fig2}(e)-\ref{fig:Fig2}(g).  Second, for the latter, the Fermi loops wind around both the median and longitude of the surface BZ torus (the 2D surface BZs formed by two $k_i$), as shown in the lower panels of Figs. \ref{fig:Fig2}(e)-\ref{fig:Fig2}(g). Moreover, the winding numbers and winding directions of the Fermi loops can be fully determined by the Chern vector. Take the $(\bar{1}00)$ surface as an example, the Chern vector components $c_2=-1$ and $c_3=1$ imply that the Fermi loop wraps along both $k_2$ and $k_3$ axis for one cycle but with opposite directions, forming a $T(-1,1)$ [or equivalently, $T(1,-1)$] torus knot. Similarly, $c_1=-1$ and $c_3=1$ leads to a $T(1,-1)$ torus knot on the $(0\bar{1}0)$ surface, and $c_1=-1$ and $c_2=-1$ results in a $T(1,1)$ torus knot on the $(001)$ surface, as shown in the lower panels of Figs. \ref{fig:Fig2}(e)-\ref{fig:Fig2}(g), respectively. 

\begin{figure}[t]
	\centering
	\includegraphics{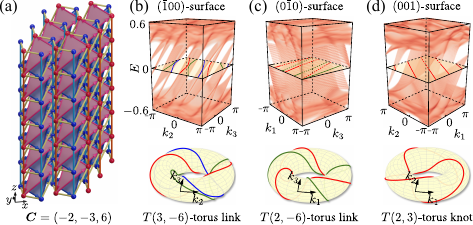}
	\caption{3D Chern insulating phases with arbitrary Chern vectors $\bm{C}=(-N_1,-N_2,N_3)$. (a) The tight-binding configuration of a $2\times 3\times 6$ supercell with a Chern vector  $\bm{C}=(-2,-3,6)$. (b)-(d) Calculated surface dispersions (upper panel) and zero-energy surface DOS (lower panel) on the $(\bar{1}00), (0\bar{1}0)$ and $(001)$ surfaces of the 3D Chern insulator shown in (a). The zero-energy Fermi loops (red, green, and blue curved lines) in (b)-(d) form distinct torus knots or torus links on different surface BZs whose characteristic integers are determined by the Chern vector. The tight-binding parameters are $t_1=1,t_2=-0.2,t_a=-t_b=0.2,M=0,\phi=\pi/2$.
	}
	\label{fig:Fig3}
\end{figure}

Now we study the propagation of chiral surface states on the acoustic 3D Chern insulator with a Chern vector $\bm{C}=(-1,-1,1)$, as shown in Fig. \ref{fig:Fig2}(h). A point source (yellow star) is placed on the  $(0\bar{1}0)$ surface to excite the chiral surface states. It can be seen that the chiral surface states propagate unidirectionally (yellow arrows) on three different adjacent surfaces. More interestingly, we find that nonreciprocal and reflectionless topological negative refraction of chiral surface states occurs at the edges  between any two adjacent surfaces, which can be explained by the iso-frequency contour analysis on two adjacent surfaces  \cite{supp}. In fact, the direction of group velocity $\bm{v}_g$ of the chiral surface states can be identified approximately as in the direction of $\bm{n}\times\bm{C}$, where $\bm{n}$ is the normal vector of the surface. A graphical illustration of this rule is provided in the Supplemental Material  \cite{supp}, which predicts the propagation and refraction direction of the chiral surface states, matching well with the simulated results in Fig. \ref{fig:Fig2}(h). It also shows that for an arbitrary surface of a 3D Chern insulator, the chiral surface states can exist as long as the Chern vector $\bm{C}$ is not in parallel with the normal vector $\bm{n}$. 

Finally, we demonstrate that 3D Chern insulating phases with arbitrary Chern vectors $\bm{C}=(-N_1,-N_2,N_3)$ can be constructed by adopting a supercell comprising $N_1 \times N_2 \times N_3$ elementary tilted unit cells shown in Fig. \ref{fig:Fig2}(a). For instance, Fig. \ref{fig:Fig3}(a) shows a 3D tight-binding model of a supercell consisting of $2\times 3 \times 6$ elementary tilted unit cells, representing a 3D Chern insulating phase with a Chern vector $\bm{C}=(-2,-3,6)$ \cite{supp}. The inter- and intra- supercell hopping strengths are staggered to ensure the selected supercell to be the irreducible primitive unit cell. Due to limited computational resources, we only perform calculation of the multi-fold supercell based on its tight-binding Hamiltonian instead of full-wave simulation. The calculated surface DOS of $(\bar{1}00), ~(0\bar{1}0)$ and $(001)$ surfaces in Figs. \ref{fig:Fig3}(b)-\ref{fig:Fig3}(d) reveal multiple chiral surface states on different surfaces due to the large Chern numbers. Moreover, we find that the winding behaviors of the Fermi loops, namely, the torus knots or links formed on the 2D surface BZs, can be greatly enriched in 3D Chern insulating phases with arbitrary Chern vectors. As shown in Fig. \ref{fig:Fig3}(b), the Fermi loops on the $(\bar{1}00)$ surface form a $T(-3,6)$ torus link because $c_2=-3$ and $c_3=6$. Since 3 and 6 are not coprime, the Fermi loops are in fact three interlinked $T(-1,2)$ [or equivalently, $T(1,-2)$] torus knots (red, green and blue twisted loops) according to the knot theory \cite{manturov_knot_2018}. Each of the three torus knots winds one cycle along $k_2$ axis and two cycles along $k_3$ axis in the opposite direction. Similarly, the Fermi loop on the  $(0\bar{1}0)$ surface forms a $T(2,-6)$ torus link [or equivalently, two interlinked $T(1,-3)$ torus knots (red and green twisted loops)] as shown in Fig. \ref{fig:Fig3}(c). The situation is different for the $(001)$ surface, since, as shown in Fig. \ref{fig:Fig3}(d), the Fermi loop forms a single $T(2,3)$ torus knot (red twisted loop) because $c_1=-2$ and $c_2=-3$. Note that these torus knots or torus links are formed in a single 3D Chern insulator, in contrast to the previously reported torus links or Hopf links formed at an interface between two different 3D Chern insulators with perpendicular Chern vectors \cite{liu_topo_2022}.  

In summary, we have extended the 3D Chern insulator phase to acoustic crystals, and have proposed a scheme to construct an acoustic 3D Chern insulator with an arbitrary Chern vector. The arbitrary Chern vector allows the emergence of surface-state torus knots or links on the surface of a single crystal, unlike the previous realization of photonic 3D Chern insulators that require a domain wall between two crystals with perpendicular Chern vectors for such phenomena. Our work thus provides a general picture of the bulk-boundary correspondence for the Chern vector. Since the acoustic 2D Chern insulator has recently been successfully realized, we envision that a similar implementation in 3D should be feasible in the near future.  

\vspace{2ex}

Z.G. acknowledges the support from the National Natural Science Foundation of China under Grant No. 62375118 and 12104211, Shenzhen Science and Technology Innovation Commission under Grant No. 20220815111105001, and SUSTech under Grant No. Y01236148 and No. Y01236248. P.P.S acknowledges the support from the National Natural Science Foundation of China under Grant No. 62220106006. L.Y. acknowledges the support from China Postdoctoral Science Foundation under Grant No. 2023M731533, Shenzhen Science and Technology Program under Grant No. RCBS20221008093326054 and SUSTech Presidential Postdoctoral Fellowship. K.X. acknowledges the support from the Cultivation of Guangdong College Students' Scientific and Technological Innovation (``Climbing Program'' Special Funds under Grant No. pdjh2023c21002). Y.M. acknowledges the support from the National Natural Science Foundation of China under Grant No. 12304484. B.Z. acknowledges the support from National Research Foundation Singapore Competitive Research Program No. NRF-CRP23-2019-0007, and Singapore Ministry of Education Academic Research Fund Tier 2 under grant No. MOE2019-T2-2-085. 
Y.Y. acknowledges the support from the Key Research and Development Program of the Ministry of Science and Technology under Grant No. 2022YFA1405200 and 2022YFA1404900, the National Natural Science Foundation of China under Grant No. 62175215, the Fundamental Research Funds for the Central Universities under Grant No. 2021FZZX001-19, and the Excellent Young Scientists Fund Program (Overseas) of China. H.C. acknowledges the support from the Key Research and Development Program of the Ministry of Science and Technology under Grant No. 2022YFA1404704 and 2022YFA1404902, the National Natural Science Foundation of China under Grant No.61975176, and the Key Research and Development Program of Zhejiang Province under Grant No.2022C01036.

\bibliography{References}

\end{document}